\documentclass[journal]{IEEEtran}
\usepackage{amsmath,bm}
\usepackage{graphics}
\usepackage{amssymb}
\usepackage{graphicx}
\usepackage{amsmath}
\usepackage{xcolor}
\usepackage{listings}
\usepackage{mdframed}
\usepackage{subcaption}
\usepackage{hyperref}
\hypersetup{
  pdftitle={},
  pdfauthor={},
    pdfnewwindow=true,      
    colorlinks=true,       
    linkcolor=blue,          
    citecolor=magenta,        
    filecolor=cyan,      
    urlcolor=teal
}

\DeclareMathOperator{\diag}{\mathbf{diag}}

\begin{document}

\title{Notch Filters without Transient Effects: A Constrained Optimization Design}





\author{Reza Sameni\thanks{R~Sameni is with the Department of Biomedical Informatics at Emory University, and the Department of Biomedical Engineering at Georgia Institute of Technology and Emory University. Email: \href{rsameni@dbmi.emory.edu}{rsameni@gatech.edu}.}}
\maketitle
\section{Background}
\label{secintro}
\par 

Transient responses are an inherent property of recursive filters due to unknown or incorrectly selected initial conditions. Well-designed stable filters are less affected by transient responses, as the impact of initial conditions diminishes over time. However, applications that require very short data acquisition periods (for example, as short as ten seconds), such as biosignals recorded and processed by wearable technologies, can be significantly impacted by transient effects. But how feasible is it to design filters without transient responses? 

We propose a well-known filter design scheme based on constrained least squares (CLS) optimization to create zero-transient effect notch filters for powerline noise cancellation. We demonstrate that this filter is equivalent to the optimal Wiener smoother in the stationary case. We also discuss its limitations in removing powerline noise with nonstationary amplitude, where a Kalman filter-based formulation can be used instead.
\section{Motivation}
\label{sec:motivation}
\par 
Autoregressive moving average models are instrumental tools in stochastic process theory and digital signal processing. At the heart of these models are linear difference equations of this form:
\begin{equation}
y_k = \sum_{m = 0}^{M} b_m x_{k-m} + \sum_{n = 1}^{N} a_n y_{k-n}
\label{eq:arma_model}
\end{equation}
where $x_k$ and $y_k$ are the inputs and outputs of the model, and $a_m$ and $b_m$ are the auto-regression and moving average coefficient sets, respectively. \textit{Finite impulse response (FIR)} models only have the moving average term. \textit{Infinite impulse response (IIR)} models have both terms, making the output dependent on the current and past inputs, in addition to the past outputs. Due to these dependencies, the output undergoes transient effects caused by unknown initial conditions ($x_{-1}, x_{-2}, \ldots$ and $y_{-1}, y_{-2}, \ldots$). In fact, if the initial conditions were known a priori (or could be estimated accurately), there would be no transient effects at the beginning.

In FIR models, transient effects exactly end after the first $M$ samples. However, in IIR models, the impact of initial conditions never completely ends. Although in well-designed stable systems the transient effects diminish over time, systems that are only marginally stable or possess narrow-band frequency responses may experience prolonged transient effects. Highly frequency-selective filters, including high quality factor (Q-factor) notch filters used for powerline noise cancellation, are one such example, which often exhibit long transient responses, resulting in \textit{ringing effects} in the output. While damping ringing effects can be tolerated in continuous filtering, applications requiring short snapshot signal acquisition and processing would greatly benefit from filtering schemes free from any transient effects.


Wearable sensors, smartwatches equipped with health monitoring sensors, and personal/mobile health monitors that record data over short periods (for example, ten to thirty seconds), and operate in real-world environments that are prone to powerline interferences are among the potential applications for \textit{transient-free} filters.

To address this need, we propose to use a constrained least squares (CLS) optimization scheme (similar to Tikhonov regularization \cite{golub1999tikhonov}), to design a transient-free powerline noise canceler. The resulting filter is linear (possessing both the homogeneity and additive properties of linear filters), but is time-variant, deviating from the linear time-invariant model in \eqref{eq:arma_model}. While the filter design scheme is rather generic, we will exemplify it for powerline noise cancellation from short-length signals. The resulting filter has no transient effect (starting from the very first sample of the input signal), does not have any delay between its input and output, acting as a non-causal fixed-interval smoother, and is very simple to implement. The resulting filter only has a single tunable parameter, and although it is time-variant, the filter coefficients are fixed and do not depend on the input signal. Therefore, for input signals of fixed-length the filter coefficients can be calculated offline and implemented as a fixed matrix multiplication.

We also investigate generalizations to long and continuous data and extensions to nonstationary powerline interference. MATLAB and Python implementations of the filter are provided as code snippets, and we also review its shortcomings for nonstationary noises.

\section{Data model}
\label{sec:data_model}
Let $s_k$ be a discrete-time signal with variance $\sigma_s^2$, and let $x_k$ represent a noisy measurement of $s_k$, impacted by the powerline interference signal $p_k$:
\begin{equation}
    x_k = s_k + p_k,\quad k = 1, \ldots, K
\label{eq:data_model}
\end{equation}
Assuming that the powerline interference is stationary, which is a valid assumption for short-lengthed measurements, $p_k$ can be modeled by a sinusoidal waveform with a fixed peak amplitude and an unknown phase:
\begin{equation}
p_k \approx A \sin\left(\omega_0 k + \theta\right)
\label{eq:sinusoidal}
\end{equation}
where $\omega_0 = 2\pi f_0/f_s$, $f_0$ is the mains frequency (typically 50\,Hz or 60\,Hz, depending on the region in which the data is collected), and $f_s$ is the sampling frequency.

Power transmission and distribution systems are designed to keep the frequency very accurate, typically with a tolerance of 0.01\% to 1\% around the mains frequency. This small frequency tolerance is due to the fact that even very small mismatches in the powerline frequency can cause damage and impose significant costs on the power network. Therefore, \eqref{eq:sinusoidal} is a reasonably accurate approximation for powerline interference signals.

Using basic trigonometry, we can show that successive samples of a sinusoidal waveform described by \eqref{eq:sinusoidal} satisfy the recursion \cite{Sameni2012}:
\begin{equation}
p_{k+1} + p_{k-1} \approx 2\cos(\omega_0) p_{k}
\label{eq:sinusoidal_recursice}
\end{equation}
where the approximation follows from the approximation in \eqref{eq:sinusoidal}, and would be accurate otherwise. Importantly, neither the powerline amplitude $A$ nor the phase $\theta$ appear in this recursion, properties that enable us to estimate and remove the powerline without knowing its peak amplitude or phase.

\section{A constrained least squares formulation for powerline estimation}
Using the data model presented in Section \ref{sec:data_model}, we seek to estimate and remove the powerline signal $p_{k}$ from noisy measurements $x_{k}$. Combining \eqref{eq:data_model} and \eqref{eq:sinusoidal_recursice}, this problem can be formulated as a CLS problem:
\begin{equation}
\begin{array}{c}
    p_l^* = \displaystyle\arg\min_{p_l} \sum_{k = 2}^{K-1}\left(p_{k+1} -2\cos(\omega_0) p_{k} + p_{k-1}\right)^2\\
    \text{s.t.}\quad \displaystyle\frac{1}{K}\sum_{k = 1}^K(x_k - p_k)^2 = \sigma_s^2
\end{array}
\label{eq:cls_sample_form}    
\end{equation}
for $l = 1,\ldots, K$. 

Defining the vectors $\mathbf{x} = [x_1, \ldots, x_K]^T$, $\mathbf{s} = [s_1, \ldots, s_K]^T$, and $\mathbf{p} = [p_1, \ldots, p_K]^T$, \eqref{eq:cls_sample_form} can be written as a Lagrangian in vector form \cite{Sameni2017}:
\begin{equation}
    \mathbf{p}^* = \arg\min_{\mathbf{p}} \|\mathbf{H} \mathbf{p}\| + \lambda \|\mathbf{x} - \mathbf{p}\| 
\label{eq:cls}    
\end{equation}
where $\mathbf{H}$ is a $(K-2) \times K$ Toeplitz matrix:
\begin{equation}
\mathbf{H} = \!\left[
\begin{array}{cccccc}
1 & -2\cos(\omega_0) & 1 & 0 & \cdots & 0 \\
0 & 1 & -2\cos(\omega_0) & 1 & \ddots & \vdots \\
\vdots & \ddots & \ddots & \ddots & \ddots & 0 \\
0 & \cdots & 0 & 1 & -2\cos(\omega_0) & 1
\end{array}
\right]
\end{equation}
This problem has a closed-form solution, similar to the well-known Tikhonov regularization, or Ridge regression \cite{golub1999tikhonov}: 
\begin{equation}
\mathbf{p}^* = (\mathbf{I} + \gamma \mathbf{H}^T \mathbf{H})^{-1}\mathbf{x}
\label{eq:opt_solution}
\end{equation}
where $\gamma = 1 / \lambda$. The matrix $\mathbf{H}^T \mathbf{H}$ is symmetric and has finite eigenvalues. Therefore, we can always set $\gamma$ such that the filtering matrix in \eqref{eq:opt_solution} is invertible. This can be shown using the \textit{singular value decomposition} (SVD) of $\mathbf{H}$:
\begin{equation}
    \mathbf{H} = \mathbf{U}\bm{\Sigma}\mathbf{V}^T
\end{equation}
where $\mathbf{U}$ is a $(K-2)\times(K-2)$ orthonormal matrix, $\mathbf{V}$ is a $K\times K$ orthonormal matrix, and $\bm{\Sigma} = \diag[\sigma_i]$ is a $(K-2)\times K $ diagonal matrix with $K-2$ non-zero singular values. Therefore, \begin{equation}
(\mathbf{I} + \gamma \mathbf{H}^T \mathbf{H})^{-1}= \mathbf{V} \diag[\epsilon_1, \ldots, \epsilon_K] \mathbf{V}^T
\label{eq:matrix_inversion}
\end{equation}
where
\begin{equation}
    \epsilon_i = \left\{\begin{array}{ll}
         \displaystyle\frac{1}{1 + \gamma\sigma_i^2} &  i = 1, \ldots, K-2\\
         0 & i = K-1, K
    \end{array}\right.
\end{equation}
which leads to:
\begin{equation}
    \begin{array}{l}
         \mathbf{p}^* = \displaystyle\mathbf{V} \diag[\epsilon_1, \ldots, \epsilon_K]\mathbf{V}^T\mathbf{x}
    \end{array}
\label{eq:opt_solution_form2}    
\end{equation}
Finally, the optimal $\mathbf{p}^*$ obtained in \eqref{eq:opt_solution_form2} can be used to calculate the powerline denoised vector $\mathbf{y} = \mathbf{x} - \mathbf{p}^*$:
\begin{equation}
    \begin{array}{l}
        \mathbf{y} = \displaystyle\mathbf{V} \diag[1-\epsilon_1, \ldots, 1-\epsilon_K]\mathbf{V}^T\mathbf{x} 
    \end{array}
\label{eq:opt_solution_output_eq}     
\end{equation}

MATLAB and Python implementations of this filter are shown in Fig.~\ref{fig:implementations}. As compared with a linear time-invariant filter implementation, the overall interpretation of the proposed powerline filter design scheme is that it uses the exact dynamics of the powerline signal, to mitigate the need for the unknown initial conditions.

The visualization of the columns of $\mathbf{V}$ is also very insightful. Each of its columns have an oscillatory behavior across the Nyquist frequency.

\begin{figure*}[tb]
    \begin{subfigure}[c]{\linewidth}\centering
\begin{mdframed}
\small{
\begin{lstlisting}[language=Matlab, basicstyle=\ttfamily, keywordstyle=\color{teal}, commentstyle=\color{brown}]
function y = notch_filter_cls(x, f0, fs, gamma)
    K = length(x); % signal length
    w0 = 2*pi*f0/fs; % angular frequency
    H = toeplitz([1; zeros(K-3, 1)], [1, -2*cos(w0), 1, zeros(1, K-3)]);
    y = x(:) - (eye(K) + gamma*(H'*H)) \ x(:); % performs least squares
end           
\end{lstlisting}}
\caption{MATLAB}
\end{mdframed}    \label{fig:matlab_implementation}
    \end{subfigure}
    \begin{subfigure}[c]{\linewidth}\centering
\begin{mdframed}
\small{
\begin{lstlisting}[language=Python, basicstyle=\ttfamily, keywordstyle=\color{teal}, commentstyle=\color{brown}]
import numpy as np
from scipy.linalg import toeplitz

def notch_filter_cls(x, f0, fs, gamma):
    K = x.shape[0]  # assuming x is a 1D column vector numpy array
    w0 = 2*np.pi*f0/fs # angular frequency
    # construct the Toeplitz matrix
    col = np.hstack(([1], np.zeros(K - 3)))
    row = np.hstack(([1, -2*np.cos(w0), 1], np.zeros(K - 3)))
    H = toeplitz(col, row)
    # performs least squares
    A = np.eye(K) + gamma * np.matmul(H.T, H)
    y = x - np.linalg.solve(A, x) 
    return y  
\end{lstlisting}
}
\caption{Python}
\end{mdframed}
    \label{fig:python_implementation}
    \end{subfigure}
\caption{MATLAB and Python implementations of the notch filter. In both implementations, $\texttt{x}$ is the input vector $\texttt{f0}$ is the notch frequency, $\texttt{fs}$ is the sampling rate, $\texttt{gamma}$ is the regularization parameter, and $\texttt{y}$ is the filtered signal.}
\label{fig:implementations}
\end{figure*}

\subsection{A Wiener filter-based perspective}
\begin{figure}[tb]
\centering
\includegraphics[trim={0 0 0 1cm},clip,width=0.99\linewidth]{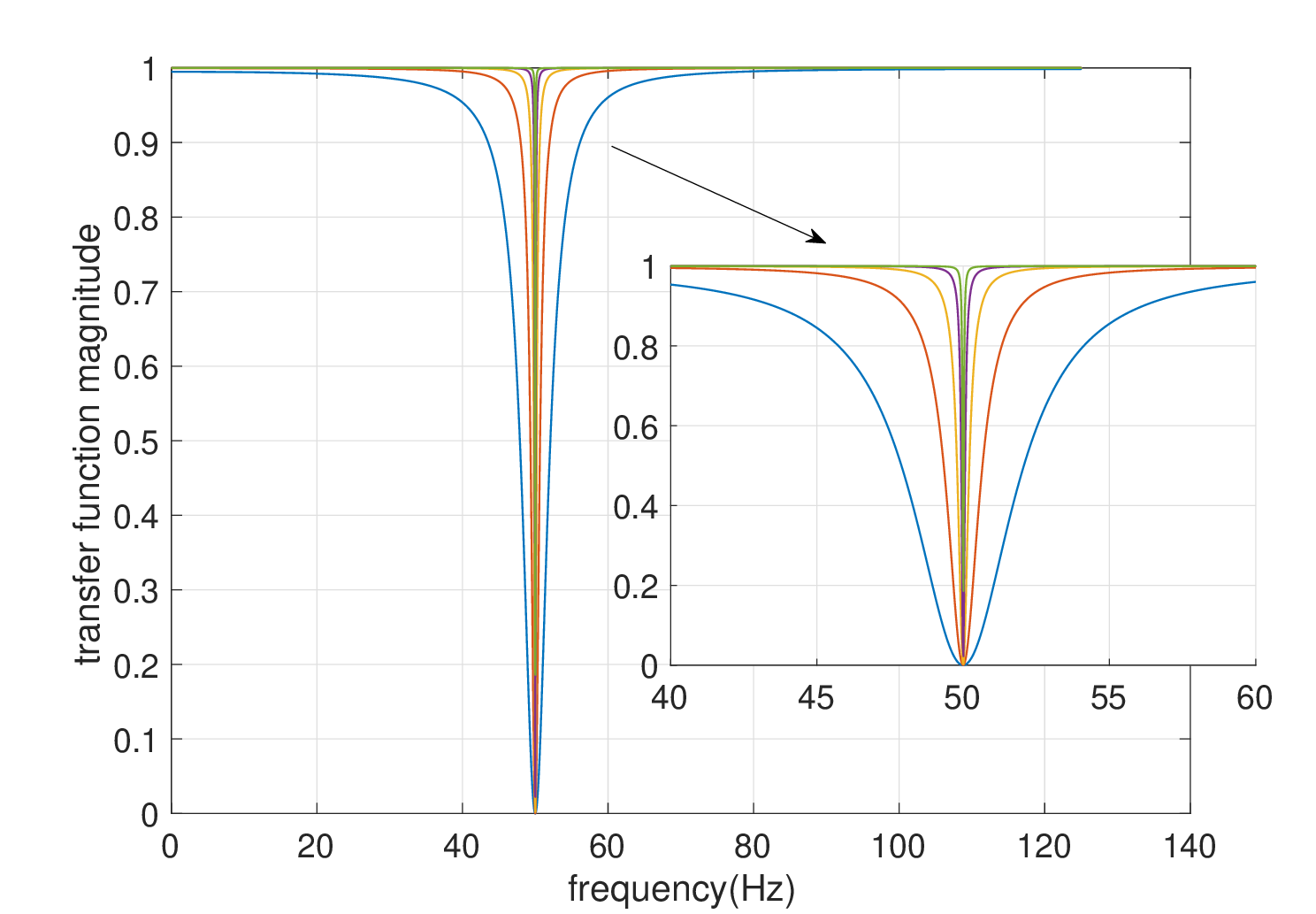}
\caption{The frequency response of the notch filter's $G(e^{j\omega})$ from \eqref{eq:tf_form_notch_filter}, for $f_0$=50\,Hz, $f_s$=250\,Hz and $\gamma$ swept between 10\textsuperscript{2} and 10\textsuperscript{6}. Higher values of $\gamma$ give higher Q-factors, resulting in sharper notch filters.}
\label{fig:freq-response}
\end{figure}

The optimization problem in \eqref{eq:cls_sample_form} can also be solved using a convolution-based IIR smoothing approach by setting the partial derivative of the cost function with respect to $p_l$ to zero. Disregarding the boundary effects and assuming an infinite length input signal, it has been shown that this approach leads to this convolution-based formulation \cite{Sameni2017}:
\begin{equation}
    (\delta_k + \gamma h_k*h_{-k})*p_k = x_k
\label{eq:conv_form}    
\end{equation}
where $h_k = [1, -2\cos(\omega_0), 1]$, $\delta_k$ is the Kronecker delta function, and $*$ denotes linear convolution. Taking the $z$-transform of \eqref{eq:conv_form}, we obtain the following transfer function for the filter:
\begin{equation}
\begin{array}{rl}
    F(z) & \!\!\!\stackrel{\Delta}{=} \displaystyle\frac{P(z)}{X(z)} = \frac{1}{1 + \gamma D(z)D(z^{-1})}
\\
& \!\!\!= \displaystyle\frac{1}{1 + \gamma[z^{-2} -4cz^{-1} + (4c^2 + 2) -4c z + z^{2}]}
\end{array}
\label{eq:tf_form}    
\end{equation}
where $P(z)$ is the Z-transform of $p_k$, $X(z)$ is the Z-transform of $x_k$, and $c\stackrel{\Delta}{=}\cos(\omega_0)$. We can show that replacing $z=e^{-j\omega}$ in \eqref{eq:tf_form}  followed by trigonometric transforms simplifies \eqref{eq:tf_form} to:
\begin{equation}
    F(e^{j\omega}) = \displaystyle \frac{1}{1 + 4\gamma[\cos(\omega) - \cos(\omega_0)]^2}
\label{eq:tf_form_frequency}    
\end{equation}
and the filter's transfer function is
\begin{equation}
\begin{array}{rl}
     G(e^{j\omega}) & \!\!\!\stackrel{\Delta}{=} \displaystyle\frac{Y(e^{j\omega})}{X(e^{j\omega})} = 1- F(e^{j\omega})\\
     & \!\!\!=\displaystyle \frac{4\gamma[\cos(\omega) - \cos(\omega_0)]^2}{1 + 4\gamma[\cos(\omega) - \cos(\omega_0)]^2}
\end{array}
\label{eq:tf_form_notch_filter}    
\end{equation}
where $Y(e^{j\omega})$ is the discrete-time Fourier transform of $y_k$. This filter notches at $\omega = \omega_0$, and for large $\gamma$, $G(e^{j\omega})\approx 1$ in all other frequencies. The realness of the filter response aligns the fact that this is a non-causal smoother with zero input-output phase/group delay. The frequency response of this filter is shown in Fig.~\ref{fig:freq-response} for $f_0$=50\,Hz, $f_s$=250\,Hz and $\gamma$ swept between 10\textsuperscript{2} and 10\textsuperscript{6}.

Importantly, $F(z)$ and $G(z)$ are non-causal (both positive and negative powers of $z$ appear in their transfer functions). If the filter was to be implemented in a linear time-invariant manner on a continuous sequence, the time domain equivalent of \eqref{eq:tf_form} would be:
\begin{equation}
p_k = x_k - \gamma p_{k-2} + 4 c \gamma p_{k-1} - (4c^2 + 2) \gamma p_k + 4 c\gamma p_{k+1} - \gamma p_{k+2}
\label{eq:time_domain_form}    
\end{equation}
which is a fourth-order noncausal filter. But using the CLS-based fornulation derived in \eqref{eq:opt_solution_form2} and \eqref{eq:opt_solution_output_eq}, we mitigate the transient effects and the need for the initial conditions in \eqref{eq:time_domain_form}.


\subsection{The nonstationary case -- a Kalman filter-based perspective}
Dynamic representations of stochastic processes commonly inspires and motivates the use of Kalman filters. As shown in \cite{Sameni2012}, using \eqref{eq:data_model} and \eqref{eq:sinusoidal_recursice}, a Kalman filter-based formulation for the notch filter problem is as follows:
\begin{equation}
\displaystyle
    \begin{array}{rl}
        \begin{bmatrix}
             p_k\\
             p_{k-1} 
        \end{bmatrix} = &
        \begin{bmatrix}
             2\cos(\omega_0) & -1\\
             1 & 0 
        \end{bmatrix}
        \begin{bmatrix}
             p_{k-1}\\
             p_{k-2} 
        \end{bmatrix} + 
        \begin{bmatrix}
             1\\
             0 
        \end{bmatrix}w_k
         \\
        x_k = & \begin{bmatrix}
             1 & 0 
        \end{bmatrix} \begin{bmatrix}
             p_k\\
             p_{k-1} 
        \end{bmatrix} + s_k  
    \end{array}
\end{equation}
where $w_k$ is a random modeling error corresponding to the approximation in \eqref{eq:sinusoidal} for minor nonstationary frequency, amplitude or phase. In this model, the signal of interest $s_k$ plays the role of an interference for the powerline signal $p_k$. Different variants of the resulting Kalman notch filters and smoothers have been developed based on this formulation \cite{Sameni2012,Warmerdam2017}.




\section{Numerical Example}
\begin{figure*}[tb]
\centering
\includegraphics[trim={0 0 0 1cm},clip,width=0.92\linewidth]{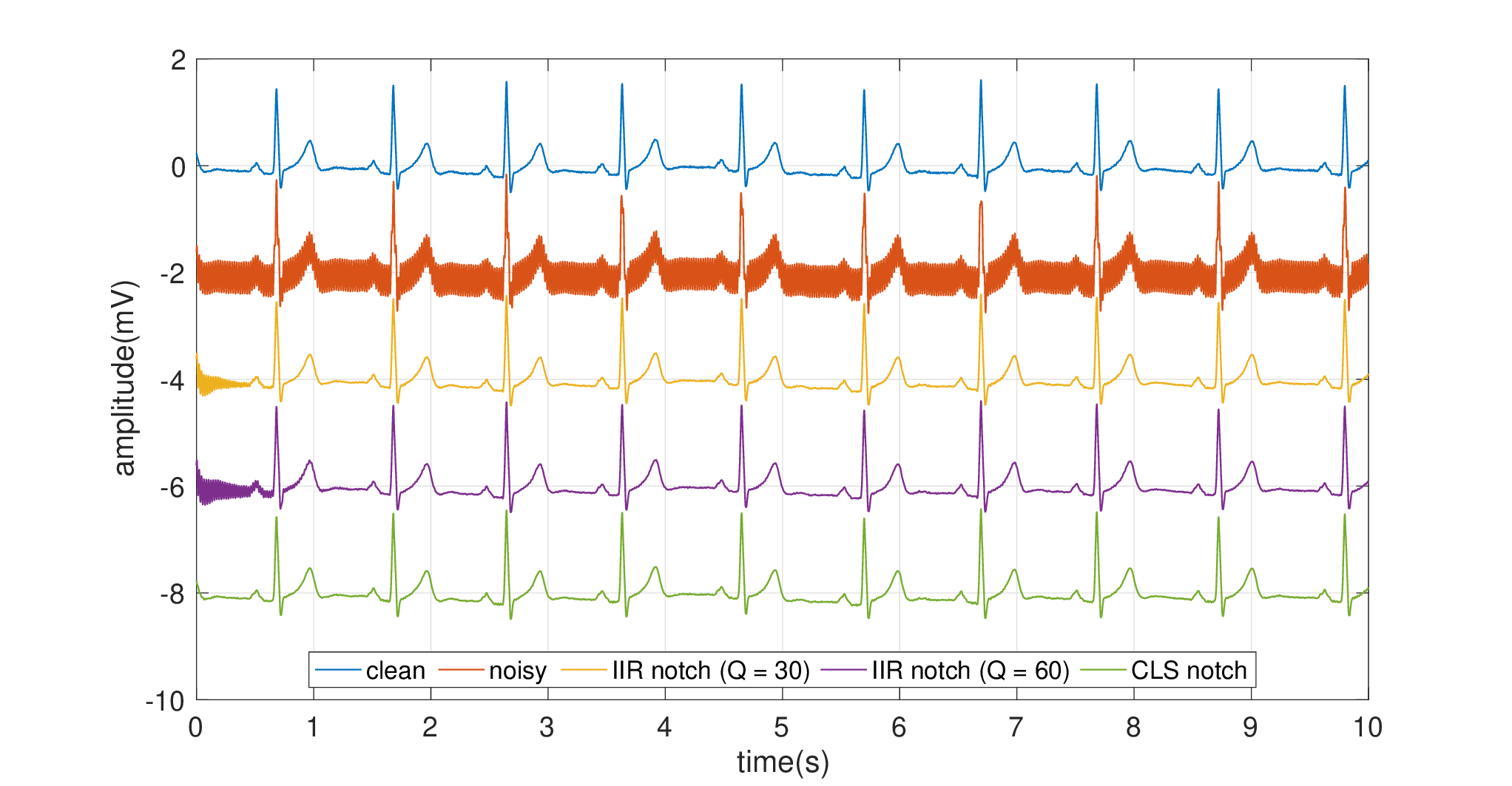}
\caption{A 10-second electrocardiogram (ECG) corrupted by additive powerline noise and filtered using standard IIR notch filters and the constrained least squares (CLS) filter from Fig.~\ref{fig:implementations}. The CLS filter does not suffer from transient effect.}
\label{fig:sample-result}
\end{figure*}

\begin{figure}[tb]
\centering
\includegraphics[trim={0 0.75cm 0 1cm},clip,width=0.95\linewidth]{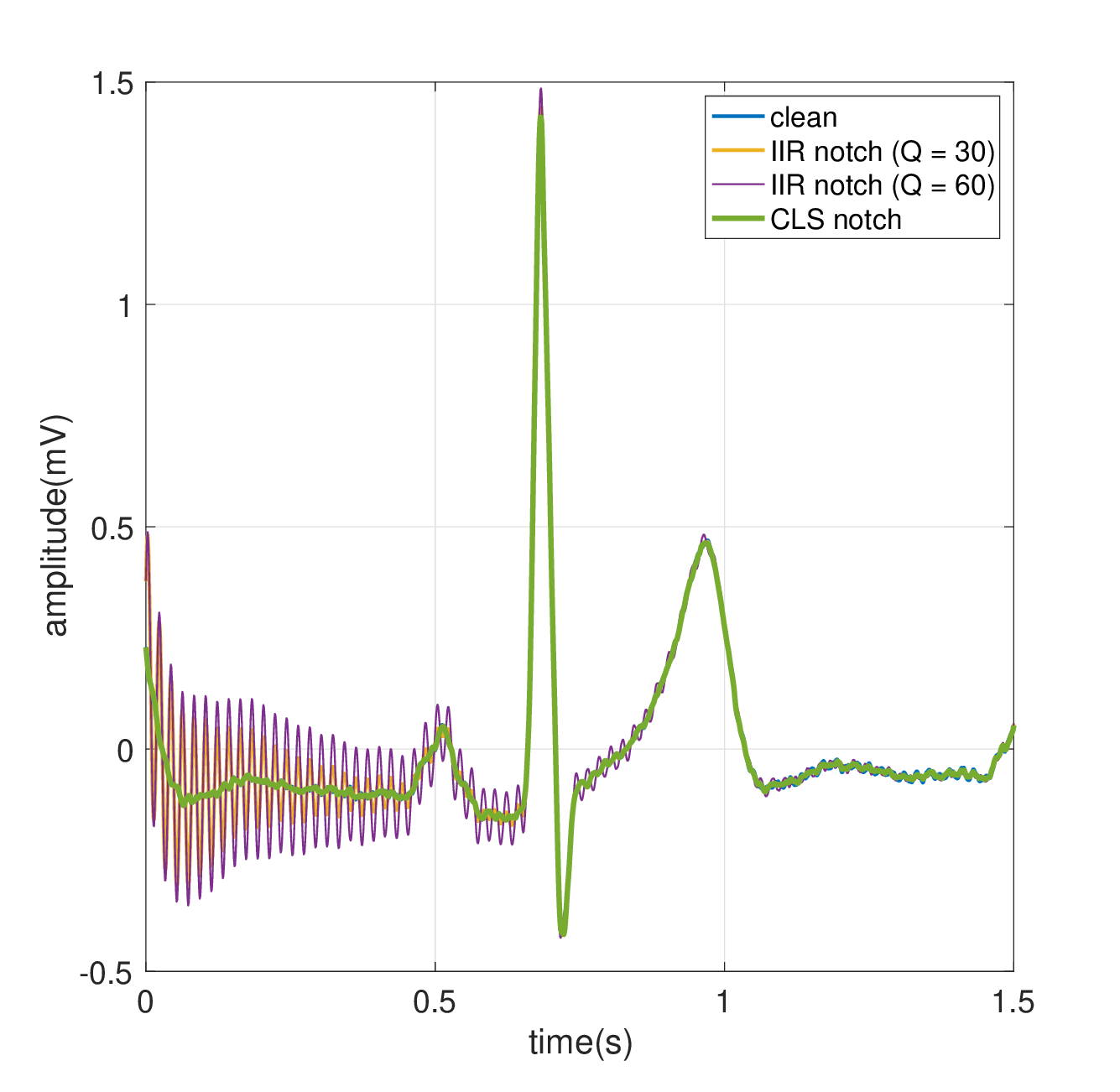}
\caption{A closer plot of the results shown in Fig.~\ref{fig:sample-result}. The CLS-based filter does not have any transient behavior.}
\label{fig:sample-result-zoomed}
\end{figure}

Fig.~\ref{fig:sample-result} demonstrates the performance of the proposed filter on an electrocardiogram (ECG) segment.
For this experiment, a 10-second ECG recorded from the PhysioNet PTB database \cite{PTB1} was selected. This dataset has been recorded in Germany, with a 50\,Hz powerline network. The selected signal did not have powerline noise. A 50\,Hz sinusoidal signal was added to the clean ECG to mimic powerline interference. The noisy signal was next filtered using MATLAB's \texttt{iirnotch} function, which is a standard second-order IIR notch filter. A similar function exists in Python (\texttt{scipy.signal.iirnotch}). We set the notch frequency to 50\,Hz and tested the filter with Q-factors of 30 and 60. The same signal was also filtered by the CLS-based notch filter using the \texttt{notch\_filter\_cls} function listed in Fig.~\ref{fig:implementations}. As we can see, while the IIR notch filters have ringing effects at the beginning of the signal, the CLS-based filter does not have any transient effects. A zoomed version of the results are shown in Fig.~\ref{fig:sample-result-zoomed} for the first 1.5\,s of the data.

The MATLAB and Python codes of the CLS notch filter, together with the test scripts are available in the open-source electrophysiological toolbox (OSET) \cite{SameniOSET314}.

\section{Limitations}
\label{seclim}
A limitation of the CLS-based filter is that the powerline noise must be stationary across the filtering window; although, this is not a significant limitation because powerline noise amplitudes are generally robust over short window lengths. However, if there is a significant jump in the powerline noise level during the filtering window, the filter will not adapt to the nonstationary variations of the noise, as the CLS scheme seeks to estimate a powerline signal with a fixed amplitude.

Another limitation of the CLS-based notch filter is the computational cost of the $K\times K$ matrix multiplication required for \eqref{eq:opt_solution}, which makes the filter practically infeasible for long records or continuous processing. Considering that the filter has no transient period and does not have any input-output lag, a practical approach to mitigate this issue is to apply the filter on small segments of the input signal (with or without overlap) and to stitch/merge the segment-wise outputs to reconstruct the entire signal. This method was used in \cite{Sameni2017} to implement a real-time version of a Tikhonov regularization filter for ECG denoising. It provides us with a fixed-interval smoother for long or continuous recordings.


\section{Conclusions}
\label{secconc}
We demonstrated the effectiveness of employing a constrained least squares (CLS)-based formulation for designing a non-causal notch filter with no transient effects. We illustrated that this filter, while analogous to a fourth-order non-causal Wiener filter, remarkably avoids any transient effects, offering a significant advantage in processing short-length signals. Its ease of implementation, necessitating merely a matrix multiplication, makes it particularly suitable for fixed-length signal applications, such as those encountered in wearable mobile health devices. Notably, by pre-calculating the matrix coefficients, the implementation of the filter can be performed by a simple matrix multiplication --- especially on devices equipped with tensor processing units (TPUs) --- enhancing its applicability on portable/mobile devices. Additionally, we outlined how this framework can be adapted for piece-wise filtering and nonstationary situations through the integration of a Kalman filter and smoothing techniques.
\bibliographystyle{IEEEtran}
\bibliography{references} 

\begin{thebibliography}{1}
\providecommand{\url}[1]{#1}
\csname url@samestyle\endcsname
\providecommand{\newblock}{\relax}
\providecommand{\bibinfo}[2]{#2}
\providecommand{\BIBentrySTDinterwordspacing}{\spaceskip=0pt\relax}
\providecommand{\BIBentryALTinterwordstretchfactor}{4}
\providecommand{\BIBentryALTinterwordspacing}{\spaceskip=\fontdimen2\font plus
\BIBentryALTinterwordstretchfactor\fontdimen3\font minus \fontdimen4\font\relax}
\providecommand{\BIBforeignlanguage}[2]{{%
\expandafter\ifx\csname l@#1\endcsname\relax
\typeout{** WARNING: IEEEtran.bst: No hyphenation pattern has been}%
\typeout{** loaded for the language `#1'. Using the pattern for}%
\typeout{** the default language instead.}%
\else
\language=\csname l@#1\endcsname
\fi
#2}}
\providecommand{\BIBdecl}{\relax}
\BIBdecl

\bibitem{golub1999tikhonov}
\BIBentryALTinterwordspacing
G.~H. Golub, P.~C. Hansen, and D.~P. O'Leary, ``Tikhonov regularization and total least squares,'' \emph{SIAM Journal on Matrix Analysis and Applications}, vol.~21, no.~1, pp. 185--194, 1999. [Online]. Available: \url{https://doi.org/10.1137/S0895479897326432}
\BIBentrySTDinterwordspacing

\bibitem{Sameni2012}
\BIBentryALTinterwordspacing
R.~Sameni, ``{A linear Kalman Notch Filter for Power-Line Interference Cancellation},'' in \emph{The 16th CSI International Symposium on Artificial Intelligence and Signal Processing (AISP 2012)}.\hskip 1em plus 0.5em minus 0.4em\relax IEEE, may 2012, pp. 604--610. [Online]. Available: \url{http://dx.doi.org/10.1109/AISP.2012.6313817}
\BIBentrySTDinterwordspacing

\bibitem{Sameni2017}
\BIBentryALTinterwordspacing
------, ``{Online filtering using piecewise smoothness priors: Application to normal and abnormal electrocardiogram denoising},'' \emph{Signal Processing}, vol. 133, pp. 52--63, apr 2017. [Online]. Available: \url{http://dx.doi.org/10.1016/j.sigpro.2016.10.019}
\BIBentrySTDinterwordspacing

\bibitem{Warmerdam2017}
\BIBentryALTinterwordspacing
G.~J.~J. Warmerdam, R.~Vullings, L.~Schmitt, J.~O. E.~H. Van~Laar, and J.~W.~M. Bergmans, ``{A Fixed-Lag Kalman Smoother to Filter Power Line Interference in Electrocardiogram Recordings},'' \emph{IEEE Transactions on Biomedical Engineering}, vol.~64, no.~8, p. 1852–1861, Aug. 2017. [Online]. Available: \url{http://dx.doi.org/10.1109/TBME.2016.2626519}
\BIBentrySTDinterwordspacing

\bibitem{PTB1}
R.~Bousseljot, D.~Kreiseler, and A.~Schnabel, ``{Nutzung der EKG-Signaldatenbank CARDIODAT der PTB uber das Internet},'' \emph{Biomedizinische Technik}, vol.~40, no.~1, pp. S317--S318, 1995.

\bibitem{SameniOSET314}
\BIBentryALTinterwordspacing
R.~Sameni, \emph{{The Open-Source Electrophysiological Toolbox (OSET), version 3.14}}, 2006--2024. [Online]. Available: \url{https://github.com/alphanumericslab/OSET}
\BIBentrySTDinterwordspacing

\end{thebibliography}
\end{document}